%% file: Lundeberg.tex
\documentclass[aps,prb,10pt,twocolumn,superscriptaddress,nobibnotes]{revtex4-1}

\usepackage{lmodern}
\usepackage[T1]{fontenc}
\usepackage[utf8]{inputenc}
\usepackage{graphicx}
\usepackage{amsmath,amssymb,amsfonts}
\usepackage[unicode=true,
            colorlinks=true,
            linkcolor=blue,
            citecolor=blue,
            urlcolor=blue]{hyperref} 
\usepackage{siunitx}
\usepackage{multirow}
\usepackage{lipsum}
\usepackage[x11names]{xcolor}

\newcommand{\domark}{%
  \vbox to 0pt{
    \kern-\dp\strutbox
    \smash{\llap{\color{red!90!black}\#\kern0.5em}}
    \vss
  }%
}

\immediate\write18{git rev-parse --short HEAD > \jobname.gitinfo }
\newcommand{\gitrev}{\InputIfFileExists{\jobname.gitinfo}{}{ref not available}}

\renewcommand{\Im}{\operatorname{Im}}

\begin{document}

\title{Tuning quantum non-local effects in graphene plasmonics}

\author{Mark B. Lundeberg}
\affiliation{ICFO --- Institut de Ciències Fotòniques, The Barcelona Institute of Science and Technology, 08860 Castelldefels (Barcelona), Spain}
\author{Yuanda Gao}
\affiliation{Department of Mechanical Engineering, Columbia University, New York, NY 10027, USA}
\author{Reza Asgari}
\affiliation{School of Physics, Institute for Research in Fundamental Sciences (IPM), 19395-5531, Tehran, Iran}
\affiliation{School of Nano Science, Institute for Research in Fundamental Sciences (IPM), 19395-5531, Tehran,
Iran} 

\author{Cheng Tan}
\affiliation{Department of Mechanical Engineering, Columbia University, New York, NY 10027, USA}
\author{Ben Van Duppen}
\affiliation{Department of Physics, University of Antwerp, Groenenborgerlaan 171, B-2020 Antwerp, Belgium}
\author{Marta Autore}
\affiliation{CIC nanoGUNE, E-20018, Donostia-San Sebastián, Spain}
\author{Pablo Alonso-González}
\affiliation{CIC nanoGUNE, E-20018, Donostia-San Sebastián, Spain}
\affiliation{Departamento de Física, Universidad de Oviedo, Oviedo 33007, Spain} 
\author{Achim Woessner}
\affiliation{ICFO --- Institut de Ciències Fotòniques, The Barcelona Institute of Science and Technology, 08860 Castelldefels (Barcelona), Spain}
\author{Kenji Watanabe}
\affiliation{National Institute for Materials Science, 1-1 Namiki, Tsukuba 305-0044, Japan}
\author{Takashi Taniguchi}
\affiliation{National Institute for Materials Science, 1-1 Namiki, Tsukuba 305-0044, Japan}
\author{Rainer Hillenbrand}
\affiliation{CIC nanoGUNE and EHU/UPV, E-20018, Donostia-San Sebastián, Spain}
\affiliation{IKERBASQUE, Basque Foundation for Science, 48011 Bilbao, Spain}
\author{James Hone}
\affiliation{Department of Mechanical Engineering, Columbia University, New York, NY 10027, USA}
\author{Marco Polini}
\email{Marco.Polini@iit.it}
\affiliation{Istituto Italiano di Tecnologia, Graphene Labs, Via Morego 30, I-16163 Genova, Italy}
\author{Frank H. L. Koppens}
\email{frank.koppens@icfo.eu}
\affiliation{ICFO --- Institut de Ciències Fotòniques, The Barcelona Institute of Science and Technology, 08860 Castelldefels (Barcelona), Spain}
\affiliation{ICREA --- Institució Catalana de Recerça i Estudis Avancats, Barcelona, Spain.}


\begin{abstract}
The response of an electron system to electromagnetic fields with sharp spatial variations is strongly dependent on quantum electronic properties, even in ambient conditions, but difficult to access experimentally.
We use propagating graphene plasmons, together with an engineered dielectric--metallic environment, 
to probe the graphene electron liquid
and unveil its detailed electronic response at short wavelengths.
The near-field imaging experiments  reveal a parameter-free match with the full theoretical quantum description of the massless Dirac electron gas, in which we identify three types of quantum effects as keys to understanding the experimental response of graphene to short-ranged terahertz electric fields.
The first type is of single-particle nature and is related to shape deformations of the Fermi surface during a plasmon oscillation.
The second and third types are a many-body effect controlled by the inertia and compressibility of the interacting electron liquid in graphene.
We demonstrate how, in principle, our experimental approach can determine the full spatiotemporal response of an electron system.
\end{abstract}

\maketitle



The quantum physics of electron systems involves complex short-distance interactions and motions that depend sensitively on electron correlations and Fermi surface deformations.\cite{PinesNozieres,GiulianiVignale}
These are often considered irrelevant in optical and transport measurements, which probe the response to electrical fields with long length scales.
When free electron systems are driven by electric fields varying rapidly in both time and space, however, the response pattern in dynamical current reveals these complex short-range effects. This aspect of electron response, known as non-locality or spatial dispersion in conductivity, arises due to the internal spreading of energy via the moving electrons.
Even in ambient conditions (as Fermi liquid parameters depend on temperature weakly\cite{PinesNozieres,GiulianiVignale}), the spatial dispersion in an electron system retains a detailed connection to Fermi-surface and electron-electron correlation effects, 
and hence it provides a unique window into quantum theories of electron systems without requiring extremes of low temperature or high magnetic field.
Unfortunately, these quantum regimes cannot be accessed by standard optical and transport probes.

Plasmons---electric waves resulting from an inertial electron conductivity combined with electric restoring forces---can act as a carrier of the spatiotemporal electric fields necessary to probe non-locality.
All systems exhibit non-local effects for plasmon wavelengths approaching the electronic Fermi wavelength $\lambda_{\rm F}$,
 which has been confirmed in experimental studies of metals and semiconductor two-dimensional (2D) electron gases.\cite{vomFelde1989,Hirjibehedin2002,Ciraci2012}
Such experiments have however led to challenges in quantitative interpretation, due to strong interactions 
that go beyond standard (e.g. random phase approximation) theoretical treatments,\cite{vomFelde1989,Hirjibehedin2002} and possible complications by edge effects and tunneling.\cite{Ciraci2012,Scholl2012,Tame2013,Li2013,Zhu2016}

In graphene, it is possible to access a different, velocity-based, type of non-locality due to the ability to tune its plasmon phase velocity to low values, close to its Fermi velocity of $v_{\rm F} \approx c/300$, where $c$ is the speed of light in vacuum.
Here, we exploit heterostructures of high-quality graphene, hexagonal Boron Nitride (h-BN), 
and nearby metal\cite{Alonso-Gonzalez2016} to confine the plasmons vertically  down to 5 nm (see insets of Fig.~\ref{fig:1}A), and as a consequence slow down the propagation velocity, as illustrated in Fig.~\ref{fig:1}A. 
In addition, by tuning to relatively low carrier density, as accessible in high-quality graphene, we can further slow plasmons down to about $c/250$, approaching the Fermi velocity, such that non-local effects become very significant. 
We probe propagating plasmons by exploiting the near-field optical scattering scanning probe technique \cite{Fei2012,Chen2012} for frequencies as low as a few THz \cite{Alonso-Gonzalez2016} to avoid interband plasmon losses. Importantly, our technique allows a very accurate determination of the plasmon wavelength, independent of edge effects, which enables us to experimentally determine the non-local dynamical conductivity of graphene, $\sigma(\omega,q)$, a function of angular frequency $\omega$ and wavevector $q$.

To illustrate the tunability of the non-local effects, we consider the effect of the environment on the graphene plasmon properties. In general, graphene plasmons occur for $\omega,q$ values that satisfy the self-oscillation condition (see Methods)
\begin{align}
\sigma(\omega,q) - \frac{i\omega}{q^2} C(\omega,q) = 0
,
\label{eq:1}
\end{align}
where $C(\omega,q)$ is the dynamical capacitance of the environment around the graphene.

For graphene--dielectric plasmons, the small capacitance values ($C = 2\varepsilon q$, for permittivity $\varepsilon$) mean that non-local regimes are only accessible for short plasmon wavelengths on the order of the Fermi wavelength.\cite{Wunsch2006,Falkovsky2007,Hwang2007,Principi2009}
In such experiments the high observed phase velocities, typically more than twice $v_{\rm F}$, have given negligible non-local effects.\cite{Woessner2015}
The addition of a metal film at distance $d$ from the graphene increases capacitance up to $C \approx \varepsilon/d$.
As a consequence of Eq.~\eqref{eq:1} and as shown in Fig.~\ref{fig:1}B this drives the plasmon to larger values of $q$, 
and thus smaller velocity. As this capacitive effect also morphs the plasmon from the graphene--dielectric unscreened dispersion 
$\omega \propto \sqrt{q}$ to a linear dispersion $\omega \propto q$, we take the plasmon phase velocity $v_{\rm p} = \omega/q$ as the key parameter characterizing plasmons in this system.\cite{Ryzhii2006,Alonso-Gonzalez2016}. 

The graphene--metal system has two tuning parameters: the separation $d$ fixed at fabrication, which controls capacitance $C$, and the graphene sheet carrier density $n_{\rm s}$, which gives {\em in-situ} control of conductivity $\sigma$. A birds-eye view of this combined tunability is shown in Fig.~\ref{fig:1}C, which shows that the combination of the two tuning parameters $n_{\rm s}$ and $d$ allows the plasmon velocity $v_{\rm p}$ to be lowered to small values near the Fermi velocity $v_{\rm F}\approx \SI{1.0e6}{m/s}$.
Figure~\ref{fig:1}C also shows how the local approximation begins to break down in this regime (with $> 10$\% error for $v_{\rm p}/v_{\rm F} < 2$) as non-local effects become significant, due to reasons that will be elaborated below.

Our experimental realization of this concept is depicted in Fig.~\ref{fig:2}.
We encapsulated graphene in the h-BN dielectric, a combination which has shown the highest-quality graphene plasmons to date,\cite{Woessner2015} and placed it on top of an AuPd metal layer.
Three devices of this form were created, with distinct graphene--metal separation $d$ controlled by the chosen thickness of the bottom h-BN layer (further fabrication details in Methods), and contacted with Au electrodes.\cite{Wang2013}
In order to visualize plasmons and characterize their propagation, we used a scattering-type near-field optical microscope in photocurrent mode.\cite{Woessner2016,Lundeberg2016,Alonso-Gonzalez2016}
We operated the microscope in the Terahertz (THz) frequency range (we chose a laser frequency of $\tfrac{\omega}{2\pi} = 3.11$~THz), as this allowed us to probe down to the lowest plasmon velocities while respecting the minimum-wavelength limitations of the near-field probe.\cite{Fei2011}
In order to detect the photocurrent, we left a narrow split in the AuPd metal layer (Fig.~\ref{fig:2}A, inset) so that a graphene p-n junction could be formed by applying distinct gate voltages $V_{\rm L}$ and $V_{\rm R}$.
As a second purpose, this sharp split also served as a launching edge for graphene plasmons. 

Plasmons appear experimentally as characteristic interference fringes in the dependence of photocurrent on tip position (Fig.~\ref{fig:2}B), due to tip--plasmon interference that modulates the absorbed power.\cite{Lundeberg2016,Alonso-Gonzalez2016}
Besides the edge-reflection fringes examined in earlier works, we also observed fringes associated with plasmons launching from the split in the AuPd gate, particularly in the thinner-$d$ devices.
In either case, the fringes allowed a determination of the plasmon wavelength $\lambda = \tfrac{2\pi}{q}$, by fitting to the photocurrent with an appropriately subtracted background (Fig.~\ref{fig:2}B; see Methods for fitting details).
This, in combination with the known excitation frequency directly yields the plasmon phase velocity $v_{\rm p}$.

The main results are shown in Fig.~\ref{fig:3}.
In each of the three devices, we extracted the plasmon phase velocity from many scanning photocurrent maps, each taken with a different gate voltage.
The data have been collated into a common form by converting gate voltage to carrier density $n_{\rm s}$ (see Methods), allowing a direct comparison with theory.
Qualitatively and consistent with the map in Fig.~\ref{fig:1}C, the smallest plasmon velocities are seen for the smallest $n_{\rm s}$ and $d$.
In Fig.~\ref{fig:3} we compare the experimental $v_{\rm p}$ values to two theories: the local approximation theory (dashed curve) shows a significant discrepancy with the data, whereas the full non-local theory (shaded curve) shows excellent agreement without any fitting parameters.
Note that the local approximation predicts plasmon velocities falling below $v_{\rm F}$ for the 5.5~nm device, in contrast to the full theory which is forbidden from this region (for reasons explained below).
We now proceed to describe this full non-local theory.

Figure~\ref{fig:4}A schematically depicts the three layers of our non-local theory, 
based on dominant effects known from electron liquid theory.\cite{PinesNozieres,GiulianiVignale} 
Including all non-local corrections, the conductivity takes the following convenient form (for frequency and wavevector below Fermi values, as in this experiment)
\begin{align}
\sigma(\omega,q) = \frac{e^2 v_{\rm F} \sqrt{|n_{\rm s}|}}{\sqrt{\pi}\hbar}\frac{i }{\omega} f\bigg(\frac{v_{\rm F} q}{\omega}\bigg)
,
\label{eq:2}
\end{align}
where $f(z)$ is a dimensionless function that describes the non-local response,
\begin{align}
f(z) &= \frac{2}{z^2}\Big(\frac{1}{(1 - z^2)^{-1/2} - 1} + \delta \Big)^{-1}
.
\label{eq:3}
\end{align}
Using this functional form we can gradually introduce the different layers of non-local response, which are plotted in Fig.~\ref{fig:4}B.
Note that the local approximation consists of ignoring the $q$-dependence (which amounts to setting $f(z) = 1$), yielding a Drude response $\sigma \propto i/\omega$ from Eq.~\eqref{eq:2}.

The first layer of non-local response is to consider the response of non-interacting electrons\cite{Ryzhii2006,Wunsch2006,Falkovsky2007,Hwang2007,Principi2009} (via random phase approximation---RPA), which is the case of Eq.~\eqref{eq:3} with $\delta = 0$.
In the RPA, conductivity $\sigma$ increases with $q$, due to the change in Fermi surface deformations.
This is closely related to Landau--Bohm--Gross dispersion\cite{Landau1946,Bohm1949} in classical plasma physics: some of the electrons, those travelling with a velocity that nearly matches $v_{\rm p}$, can interact longer with each passing wavefront and thereby provide enhanced reponse (Fig.~\ref{fig:4}A).
Classically, this non-local dispersion would come along with Landau damping due to fast thermal electrons that fully match the plasma velocity and dissipate energy; this does not occur in a quantum degenerate system due to the narrowly-distributed electron velocity (the Fermi velocity), which instead yields a divergent intraband contribution to the conductivity as $q \rightarrow \omega/v_{\rm F}$ (i.e., $z\rightarrow 1$), and no Landau damping before this point.
This divergence results in the non-local plasmon velocity never falling below the Fermi velocity (as can be seen in Fig.~\ref{fig:3}), in striking contrast to the prediction of a local approximation.

The second and third layers of our non-local theory involve microscopic electron-electron interactions (many-body effects).
We have calculated these many-body corrections fully consistently, including the realistic screening by capacitance $C(\omega,q)$ (see Methods).
The major many-body effect is renormalization of band structure, which in graphene amounts to an increase in Fermi velocity (Fig.~\ref{fig:4}B).
While the value $v_{\rm F} = \SI{1.0e6}{m/s}$ is nominally assumed in graphene plasmon studies, the Fermi velocity actually varies logarithmically with carrier density, from its bare value of $\SI{0.85e6}{m/s}$ up to as much as $\SI{3.0e6}{m/s}$ for very low carrier densities.
\cite{Abedinpour2011,elias2011dirac,Orlita2012,DasSarma2013,Levitov2013} 
Since our experiments enter a regime of relatively low densities, it is crucial to include this $n_{\rm s}$-dependent velocity renormalization. 
The secondary many-body effect has to do with electron liquid correlations, which produce a Pauli-Coulomb hole\cite{PinesNozieres,GiulianiVignale} around each reference electron (Fig.~\ref{fig:4}C).
We capture this by including a local field factor $G(\omega,q)$, via $\sigma^{-1} = \sigma_{\rm RPA}^{-1} + q^2 G/(i\omega C)$, that forces consistency between the dynamic response and the isothermal compressibility.\cite{GiulianiVignale}
As described in the Methods, in our experimental regime this ultimately introduces a factor $\delta = 1 - \tfrac{\kappa_0}{\kappa}$ into Eq.~\eqref{eq:3}, where $\kappa_0$ is the RPA compressibility, and $\kappa$ is the proper isothermal compressibility \cite{Orlita2012,Hwang2007,Yu2013}.

Figure~\ref{fig:4}B shows how we can isolate the graphene conductivity function $\sigma(\omega,q)$ to directly observe the non-locality.
This is possible due to Eq.~\eqref{eq:1}, which implies that a determination of the plasmon wavevector, $q_{\rm p} = q_{\rm p}(\omega)$, produces a measurement of the dynamical conductivity at that wavevector:
$\Im \sigma(q_{\rm p},\omega) = (\omega/q_{\rm p}^2) C(q_{\rm p},\omega)$.
We are able to exactly calculate $C(\omega,q)$ from Maxwell equations (see Methods), and hence this approach of introducing variable $d$ (causing variable $C$ and variable $q_{\rm p}$) allows us to map out $\Im \sigma(q_{\rm p})$ as shown in Fig.~4.
Each device (differing in $d$) thus provides a distinct probe of 
the functional dependence of the conductivity on wavevector $q$ under otherwise-identical parameters ($\omega, n_{\rm s}$).
It can be seen in Fig.~\ref{fig:4}B that our data are only matched by theory after taking into account all three layers of quantum corrections, and that the measured conductivity of graphene shows significant departures from the local theory (a horizontal line).


The recipe set forth in this work can be transferred to probe other electron systems with exotic physical properties. 
Not only does this technique reveal the collective excitation (plasmon), but we have also shown how one may isolate the electronic response from its environment, quantitatively mapping out the underlying response function (non-local conductivity) as a function of wavelength.
This kind of spatial spectroscopy forms a valuable counterpart to the traditional temporal (frequency) spectroscopy, and the marriage of these two approaches into a precision spatiotemporal spectroscopy---a full determination of $\sigma(\omega,q)$---would provide an unprecedented window into electron physics.
This may allow a greatly enriched understanding of electron correlation physics such as those underlying fractional quantum hall effects (e.g., magneto-rotons\cite{Girvin1986,Kukushkin2009,Haldane2011}) and the binding mechanism in superconductors,\cite{Kirzhnits1973} 
as well as probing the non-locality of Fermi-surface deformations in unusual band structures 
(e.g., Weyl fermions\cite{xu2015discovery,Hofmann2016,Pellegrino2015,soluyanov2015type,Song2017}). 


%
\begin{acknowledgments}
{ \small

F.H.L.K., M.P., and R.H. acknowledge support by the European Union Seventh Framework Programme under grant agreement no.~696656 Graphene Flagship. M.P. acknowledges support by Fondazione Istituto Italiano di Tecnologia. F.H.L.K. acknowledges financial support from the from the European Union Seventh Framework Programme under the ERC starting grant (307806, CarbonLight) and project GRASP (FP7-ICT-2013-613024-GRASP). F.H.L.K. acknowledges support from the Spanish Ministry of Economy and Competitiveness, through the “Severo Ochoa” Programme for Centres of Excellence in R\&D (SEV-2015-0522), support by Fundacio Cellex Barcelona, CERCA Programme / Generalitat de Catalunya,  the Mineco grants Ramón y Cajal (RYC-2012-12281), Plan Nacional (FIS2013-47161-P and FIS2014-59639-JIN), and support from the Government of Catalonia trough the SGR grant (2014-SGR-1535).

P.A.-G. and R.H. acknowledge support from the European Union through ERC starting grant (TERATOMO grant no. 258461) and the Spanish Ministry of Economy and Competitiveness (national project MAT2012-36580). Y.G., C.T., and J.H. acknowledge support from the US Office of Naval Research N00014-13-1-0662.
C.T.\ was supported under contract FA9550-11-C-0028 and awarded by the Department of Defense, Air Force Office of Scientific Research, National Defense Science and Engineering Graduate (NDSEG) Fellowship, 32 CFR 168a. This research used resources of the Center for Functional Nanomaterials, which is a U.S. DOE Office of Science Facility, at Brookhaven National Laboratory under Contract No. DE-SC0012704. B.v.D
acknowledges support from the Flemish Science Foundation (FWO-Vl) by a post-doctoral Fellowship. P.A-G, acknowledges financial support from the national project FIS2014-60195-JIN and the ERC starting grant 715496, 2DNANOPTICA. This work used open source software (www.python.org, www.matplotlib.org, www.povray.org).
} \end{acknowledgments}
%

%
\section*{Competing financial interests}
{ \small
R.H. is co-founder of Neaspec GmbH, a company producing scattering-type scanning near-field optical microscope systems such as the ones used in this study.
All other authors declare no competing financial interests.
}


\clearpage
\input{fig1.tex}
\input{fig2.tex}
\clearpage

\input{fig3.tex}
\input{fig4.tex}

\end{document}

%% file: fig1.tex
\begin{figure}[!p]
\raggedright
\includegraphics[scale=0.32]{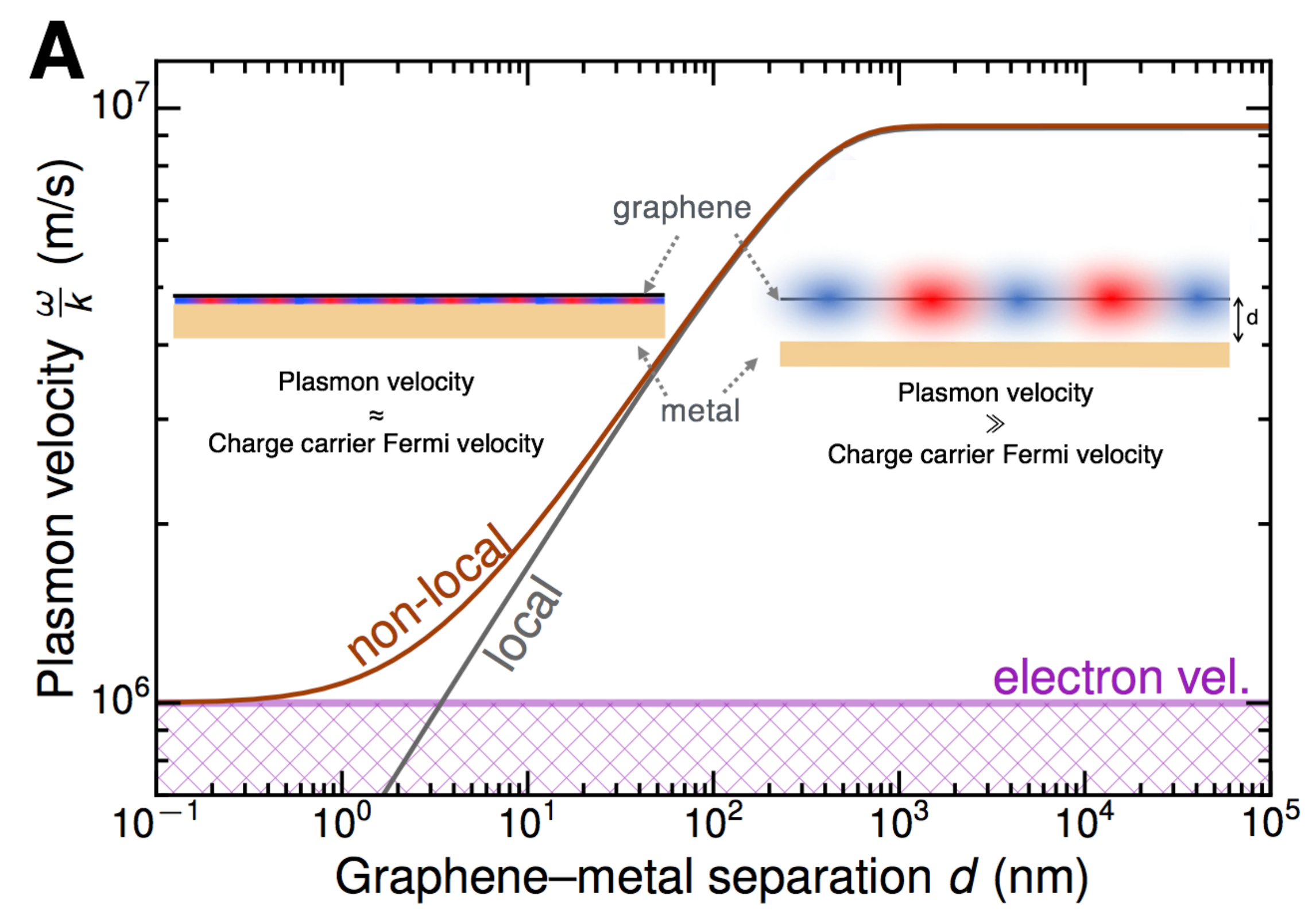}
\includegraphics[scale=1.3]{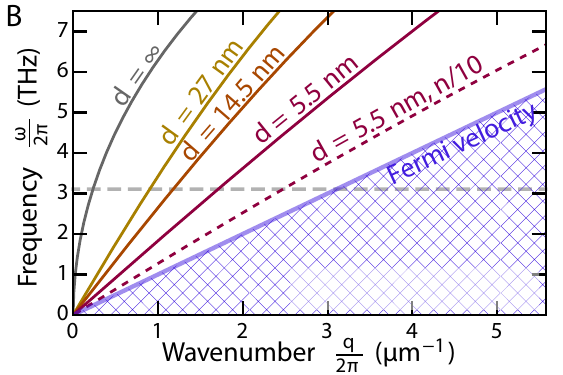}
\includegraphics[scale=1.3]{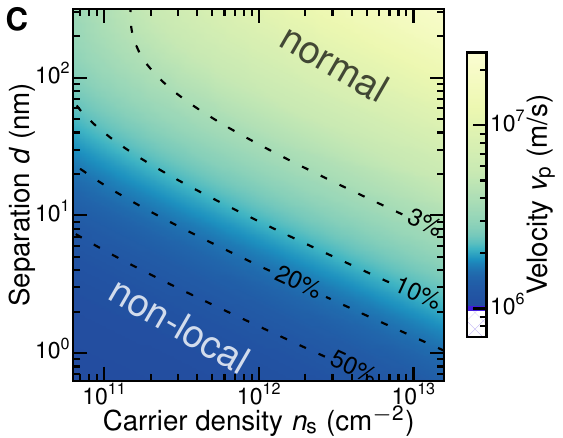}
 \caption{
 {\bf Concept.}
 ({\bf A}) Effect on graphene plasmon velocity from changing separation $d$ in the graphene--metal system. Insets show plasmon electric fields for large or small separation $d$.
 ({\bf B}) Frequency--wavenumber dispersion of plasmon at various $d$; the solid lines are all computed with equal carrier density $n_{\rm s}=10^{12}~\hbox{cm}^{-2}$, whereas the dashed line shows the smallest-$d$ case with a factor 10 lower carrier density $n_{\rm s}=10^{11}~\hbox{cm}^{-2}$. Horizonal dashed gray line indicates frequency for which the experiment has been performed.
 ({\bf C}) Plasmon velocity dependence on $d$ and carrier density $n_{\rm s}$.
 Contours indicate discrepancy between local and non-local plasmon models.
 \label{fig:1}
 }
\end{figure}

%% file: fig2.tex
\begin{figure*}[p!]
%
\hspace{-0.35in}
\includegraphics[scale=1.15]{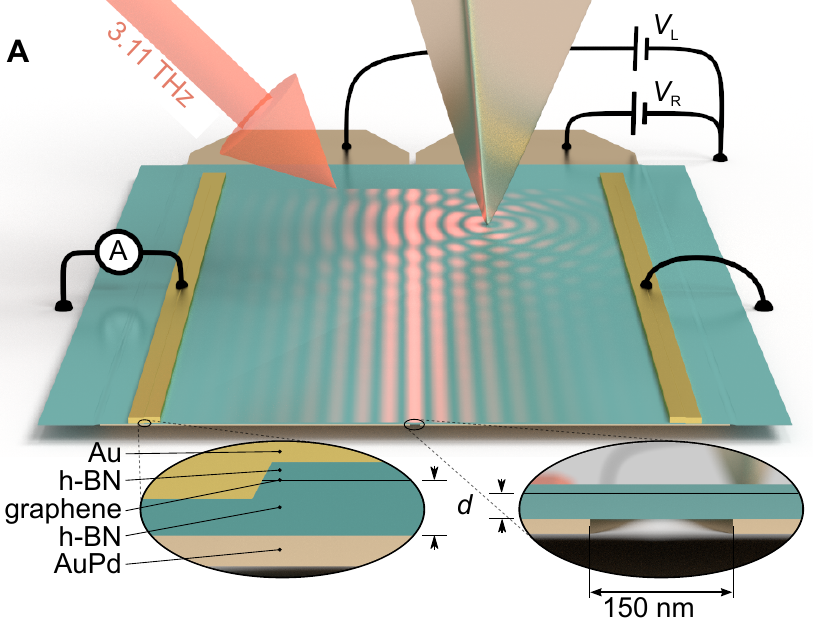}%
\includegraphics[scale=1.15]{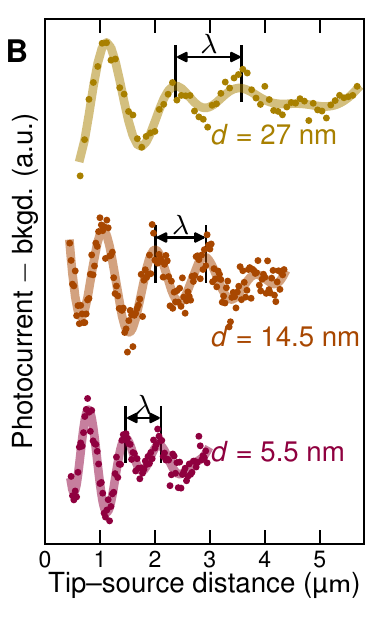}
 \caption{
 {\bf Experiment.}
 ({\bf A}) A metallized tip (inverted pyramid) scans over a graphene sheet that has been encapsulated in h-BN and placed on a split metallic film.
 Terahertz laser light illuminates the entire device, launching plasmons (orange arrows) at the tip and split.
 Gate voltages $V_L$ and $V_R$ control the electron density and the junction photocurrent sensitivity.
 ({\bf B}) Photocurrent traces in three different devices, each at $n_{\rm s} = \SI{1e12}{cm^{-2}}$, showing interference fringes used to extract the plasmon wavelength $\lambda$ (and hence velocity $v_{\rm p} = \lambda \tfrac{\omega}{2\pi}$) via the indicated fits.
 \label{fig:2}
 }
\end{figure*}

%% file: fig3.tex
\begin{figure*}[p!]
\includegraphics{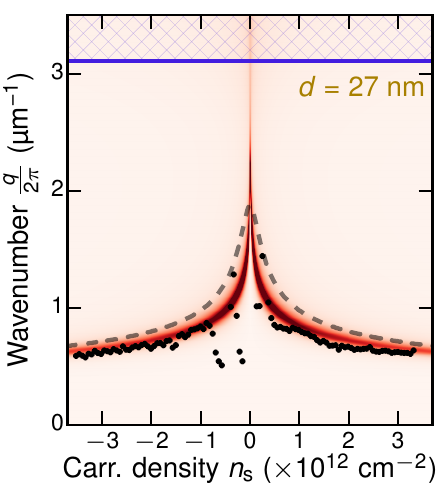}%
\includegraphics{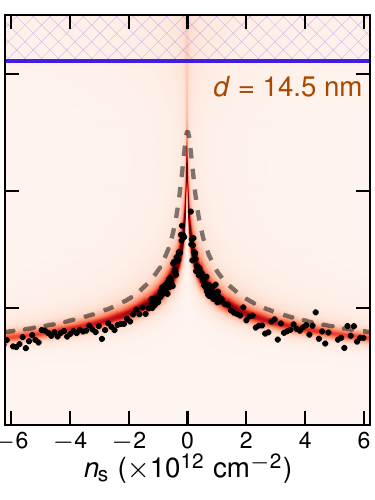}%
\includegraphics{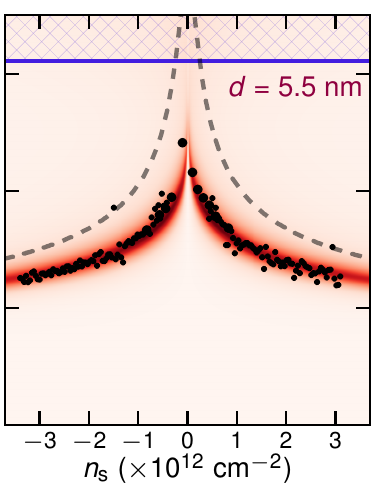}
\includegraphics{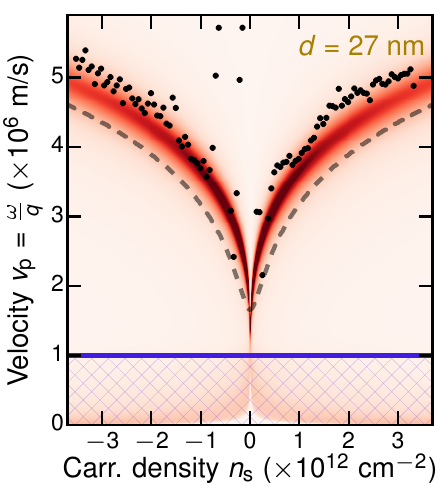}%
\includegraphics{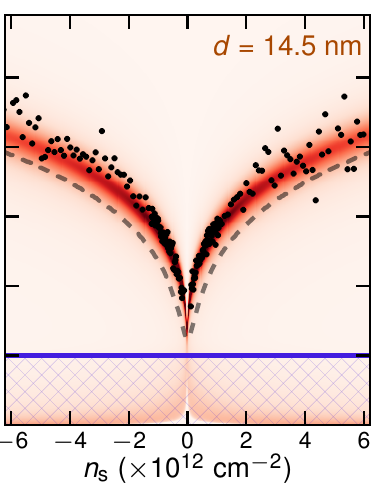}%
\includegraphics{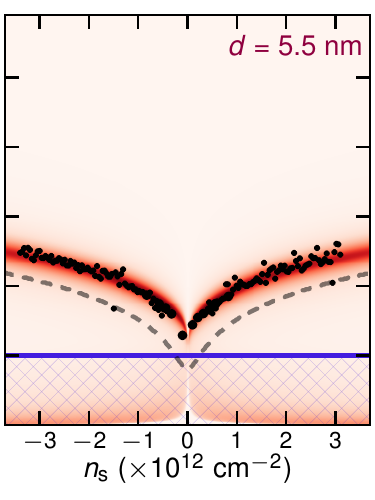}
 \caption{
 {\bf Tunable non-local effects.}
 Extracted plasmon wavelength dependence on carrier density $n_{\rm s}$, for three devices of differing separation $d$, show parameter-free agreement with the full theory (color map) and significant deviation from a local-response theory (dashed line).
 The red color map indicates the inverse of the left-hand-side of Eq.~\eqref{eq:1}, so that plasmons appear as a red peak, the width being associated with propagation distance.
 The hatched region below the solid line indicates phase velocities below $v_{\rm F}$.
 The upper and lower rows show the same data, plotted differently.
 \label{fig:3}
 }
\end{figure*}

%% file: fig4.tex
\begin{figure*}[p!]
\includegraphics[scale=1.7]{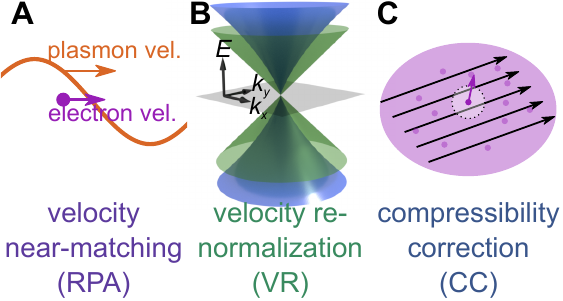}
\includegraphics[scale=1.7]{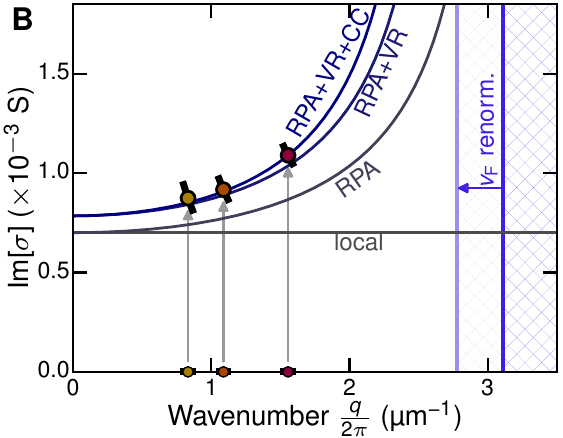}
 \caption{
 {\bf Non-local conductivity of graphene.}
 ({\bf A},{\bf B},{\bf C})
 Schematic representations of the three main mechanisms governing graphene response beyond the local approximation.
 ({\bf D})
 Experimentally-extracted $\sigma(\omega,q)$ at $n_{\rm s} = \SI{1e12}{cm^{-2}}$, compared to theoretical approximations for the interacting electron system in graphene:
 random phase approximation (RPA), with added velocity renormalization (RPA+VR), and then with compressibility correction (RPA+VR+CC); local RPA appears as a horizontal line.
 \label{fig:4}
 }
\end{figure*}